\documentclass[twocolumn,twocolappendix,trackchanges,resetfootnote]{aastex701}

\usepackage{multirow}

\shorttitle{MIRI Imaging of the SR 12 c Disk}
\shortauthors{Wu et al.}

\published{in ApJ}

\begin{document}

\title{{\large Mid-infrared Imaging of SR 12 c with JWST\slash MIRI: Grain Growth in an Evolved Circumplanetary Disk}}

\author[orcid=0000-0002-4392-1446,gname='Ya-Lin',sname='Wu']{Ya-Lin Wu}
\affiliation{Department of Physics, National Taiwan Normal University, Taipei 116, Taiwan}
\email[]{}  
\email[show]{yalinwu@ntnu.edu.tw}

\author[orcid=0000-0002-9884-9584,gname='Claire',sname='Finley']{Claire O. Finley}
\affiliation{Department of Astronomy, University of Texas at Austin, Austin, TX 78712, USA}
\affiliation{Department of Physics, University of California, Santa Barbara, Santa Barbara, CA 93106, USA}
\email[]{}  

\author[orcid=0000-0003-2649-2288,gname='Brendan',sname='Bowler']{Brendan P. Bowler}
\affiliation{Department of Physics, University of California, Santa Barbara, Santa Barbara, CA 93106, USA}
\email[]{}  

\author[gname='Ning',sname='Chen']{Ning Chen}
\affiliation{Department of Physics, National Taiwan Normal University, Taipei 116, Taiwan}
\email[]{} 

\author[orcid=0000-0003-0568-9225,gname='Yuhiko',sname='Aoyama']{Yuhiko Aoyama}
\affiliation{School of Physics and Astronomy, Sun Yat-sen University, Zhuhai 519082, People's Republic of China}
\email[]{}  

\author[orcid=0000-0002-7154-6065,gname='Gregory',sname='Herczeg']{Gregory J. Herczeg}
\affiliation{Kavli Institute for Astronomy and Astrophysics, Peking University, Beijing 100871, People's Republic of China}
\affiliation{Department of Astronomy, Peking University, Beijing 100871, People's Republic of China}
\email[]{} 

\author[orcid=0000-0001-9227-5949,gname='Catherine',sname='Espaillat']{Catherine C. Espaillat}
\affiliation{Department of Astronomy, Boston University, Boston, MA 02215, USA}
\affiliation{Institute for Astrophysical Research, Boston University, Boston, MA 02215, USA}
\email[]{}  

\author[orcid=0000-0003-0562-1511,gname='William',sname='Best']{William Best}
\affiliation{Department of Astronomy, University of Texas at Austin, Austin, TX 78712, USA}
\email[]{}  

\author[orcid=0000-0002-1483-8811,gname='Ian',sname='Czekala']{Ian Czekala}
\affiliation{School of Physics \& Astronomy, University of St Andrews, St. Andrews, KY16 9SS, UK}
\email[]{}  

\author[orcid=0000-0002-7821-0695,gname='Katherine',sname='Follette']{Katherine B. Follette}
\affiliation{Physics and Astronomy Department, Amherst College, Amherst, MA 01002, USA}
\email[]{}  

\author[orcid=0000-0001-9811-568X,gname='Adam',sname='Kraus']{Adam L. Kraus}
\affiliation{Department of Astronomy, University of Texas at Austin, Austin, TX 78712, USA}
\email[]{}  

\author[orcid=0000-0001-6301-896X,gname='Raquel',sname='Martinez']{Raquel A. Martinez}
\affiliation{Department of Physics and Biophysics, University of San Diego, San Diego, CA 92110, USA}
\email[]{}  

\author[orcid=0000-0003-1639-510X,gname='Connor',sname='Robinson']{Connor E. Robinson}
\affiliation{Division of Physics and Astronomy, Alfred University, Alfred, NY 14802, USA}
\email[]{}  

\author[orcid=0000-0001-6532-6755,gname='Quang',sname='Tran']{Quang H. Tran}
\altaffiliation{51 Pegasi b Fellow}
\affiliation{Department of Astronomy, Yale University, New Haven, CT 06511, USA}
\affiliation{Department of Astronomy, The University of Texas at Austin, Austin, TX 78712, USA}
\email[]{}  

\author[orcid=0000-0002-4479-8291,gname='Kimberly',sname='Ward-Duong']{Kimberly Ward-Duong}
\affiliation{Department of Astronomy, Smith College, Northampton, MA 01063 USA}
\email[]{} 

\author[orcid=0000-0003-2969-6040,gname='Yifan',sname='Zhou']{Yifan Zhou}
\affiliation{Department of Astronomy, University of Virginia, Charlottesville, VA 22904, USA}
\email[]{}

\begin{abstract}
The physical properties of circumplanetary disks offer insights into the formation of giant planets and their satellites. The circumbinary planetary-mass companion SR 12 c (1200 au from SR 12 AB) hosts an isolated dust disk directly detected with ALMA. We present JWST\slash MIRI imaging of SR 12 c from 5.6 to 21~\micron~and analyze its spectral energy distribution from infrared to submillimeter wavelengths. Phenomenologically, the dust disk can be described by a two-component blackbody with large contrasts in characteristic temperatures and radii. Physically motivated models with varied geometries and heating mechanisms suggest a predominantly flat, irradiation-heated disk with a smaller contribution from viscous dissipation. The disk is inferred to be compact, with an inner radius of $\sim$3 $R_{\rm Jup}$. The outer radius is only loosely constrained, with a characteristic value of $\sim$3 au. The lack of strong 10~\micron~excess emission suggests significant grain growth and dust settling. We also find that SR 12 c is broadly consistent with the empirical star--disk scaling relations, providing context for its possible formation pathway. These results demonstrate the power of combining mid-infrared and submillimeter observations with disk modeling to better understand the physical properties of circumplanetary disks.\\[7pt] 
{\it Unified Astronomy Thesaurus concepts:} Extrasolar gaseous giant planets (509); Accretion (14)\\
\end{abstract}

\section{Introduction}
Circumplanetary disks are gravitationally bound, rotationally supported structures of gas and solids surrounding giant planets. By mediating the transport of mass and angular momentum from the surrounding environment to the planet, they regulate planetary accretion rates, spin evolution, and thermal structure, and thus play a central role in giant planet formation (e.g., \citealt{D03,LM13}). Beyond their role in planetary growth, circumplanetary disks can directly influence the formation, migration, and long-term stability of regular satellites (e.g., \citealt{CW02}), offering a unique window into the early history of the solar system. Despite their importance, they remain among the least observationally constrained components of giant planet formation. Deep searches targeting gaps of circumstellar disks with the Atacama Large Millimeter\slash submillimeter Array (ALMA) have mostly yielded nondetections (e.g., \citealt{Andrews21,Facchini26}). The small number of known protoplanets also limits such searches. To date, the only compelling submillimeter detection is associated with the protoplanet PDS 70 c \citep{Isella19,B21}. 

In this context, wide-orbit planetary-mass companions offer a complementary opportunity to study circumplanetary disks in relative isolation. Observationally, about half of the young companions exhibit clear signatures of accretion and disk emission (e.g., \citealt{Bowler17,MK22}). Despite lacking continuous inflow from a circumstellar disk, their disks may share broadly similar underlying physics (e.g., viscous evolution, dust processing) with those around embedded protoplanets. Theoretically, wide-orbit companions are expected to host larger disks, since their larger Hill radii allow for more extended structures (e.g., \citealt{SB13}). However, early ALMA observations have consistently resulted in upper limits (e.g., \citealt{Bowler15,W17,W20}), implying that isolated circumplanetary disks are substantially more compact. Deep integrations are therefore essential, as demonstrated by the successful 0.88 mm continuum detection of the disk around SR 12 c \citep{W22}.

Mid-infrared observations are particularly valuable for constraining the disk structure and mineralogy of circumplanetary disks, especially with the Mid-Infrared Instrument (MIRI; \citealt{Rieke15}) on the James Webb Space Telescope (JWST). The shape of the spectral energy distribution (SED) provides insight into the disk structure and heating mechanisms, while the silicate features around 10~\micron~probe the grain composition and vertical structure of the disk. Recent studies of several planetary-mass companions suggest that their disks have undergone substantial dust evolution \citep{Cugno24,Hoch25,Malin25}. 

We have conducted a JWST/MIRI imaging survey of isolated circumplanetary disks spanning 5.6--21~\micron. Here we perform SED fitting from infrared to submillimeter wavelengths to constrain the structure and heating sources of the disk around SR 12 c, quantify the 10~\micron~excess, and discuss the potential formation scenarios. The ALMA measurement extends the SED coverage to submillimeter wavelengths and provides important constraints on the colder component of the dust disk, particularly the outer radius and total disk luminosity. The inferred disk radius can be further compared with empirical star--disk scaling relations observed for young stars and brown dwarfs, offering insights into the possible formation pathway of SR 12 c. UV-optical observations with the Hubble Space Telescope (HST) and the modeling of the accretion shock are presented in our companion paper \citep{Finley26}. Together, these studies provide the broadest wavelength coverage of a young accreting planetary-mass companion to date.

\section{Physical Properties of SR 12 c}
Discovered by \cite{K11}, SR 12 c is separated by $8\farcs7$ from the host binary SR 12 AB, or a projected separation of 1200 au using a distance of $139\pm5$ pc \citep{Gaia23, R23}. Using a revised extinction of $A_V=0.85\pm0.17$ mag \citep{Finley26}, we apply the $J$-band bolometric correction in \cite{Filippazzo15} to obtain a bolometric luminosity of log$(L/L_{\odot})=-2.82\pm0.10$ dex. Assuming a median age of 3--4 Myr for the low-extinction population toward the L1688 cluster in the Ophiuchus star-forming complex \citep{EL20}, evolutionary models of \cite{Baraffe15} yield a mass of $16\pm2~M_{\rm Jup}$. The near-infrared spectrum indicates an effective temperature of $2600\pm100$ K \citep{SM18}, corresponding to a radius of $1.9\pm0.3~R_{\rm Jup}$ and a surface gravity of log $g$ = $4.0\pm0.2$.

SR 12 c is actively accreting at a rate of $\sim$$10^{-11}~M_\odot$ yr$^{-1}$ \citep{SM18,Finley26}. Its dust disk is unresolved by ALMA, implying a size less than a few astronomical units, more compact than expected from tidal truncation \citep{W22}. The flux density of 127 $\mu$Jy corresponds to an opacity-dependent dust mass of 0.02--0.08 $M_\oplus$, which is comparable to the disk around PDS 70 c, but about 5 times lower than that of the free-floating planet OTS 44 \citep{W22}.

\section{Observations}
The Cycle 1 JWST MIRI imaging on SR 12 (GO 2311; PI: Y.-L. Wu) was performed on UT 2023 March 7 in F560W, F770W, F1000W, F1280W, F1500W, F1800W, and F2100W. We used the SUB256 subarray and the four-point dither to reduce the readout time and retain a sufficient overlapped region to image the entire system. The group time, pixel scale, and the field of view for each frame are 0.3 s, 110.5 mas, and $28\farcs3$, respectively. Table \ref{tb:phot} lists the pivot wavelength and the exposure time for each filter. The exposure time was designed to increase with wavelength from 64.7 s in F560W to 1208.9 s in F2100W to ensure that even the disk-free photosphere of SR 12 c could be detected with a signal-to-noise ratio greater than 10. We used the Stage 3 products (JWST Science Calibration Pipeline version 1.18.0), which have been processed to produce science‑ready mosaics with absolute flux calibration and corrected geometric distortion, for our analysis. All the MIRI data used in this paper can be found in the Mikulski Archive for Space Telescopes:\dataset[https://doi.org/10.17909/0h2a-k821]{https://doi.org/10.17909/0h2a-k821}.

\begin{deluxetable*}{lrrrr}
\tablecaption{Observations and Flux Densities of SR 12 c\label{tb:phot}}
\tablehead{
\colhead{Filter} & 
\colhead{$\lambda_{\rm pivot}$} & 
\colhead{$T_{\rm int}$} &
\colhead{Observed} & 
\colhead{Dereddened} \\
\colhead{} &
\colhead{(\micron)} &
\colhead{(s)} &
\multicolumn{2}{c}{($10^{-19}$ erg\,s$^{-1}$\,cm$^{-2}$\,\AA$^{-1}$)} 
}
\startdata
F560W  	&   5.635&	     64.7	& $67.22 \pm 0.68$   &  $68.27 \pm 0.72$   \\
F770W  	&   7.639&     64.7	& $35.81 \pm 0.34$   &  $36.39 \pm 0.36$   \\
F1000W 	&   9.953&     64.7 	& $22.16 \pm 0.20$   &  $23.52 \pm 0.35$   \\
F1280W 	& 12.810&   124.6	& $12.19 \pm 0.23$   &  $12.45 \pm 0.24$  \\
F1500W 	& 15.064&	   310.3	& $8.89 \pm 0.24$   &    $9.04 \pm 0.25$  \\
F1800W 	& 17.983&	   609.8	& $6.21 \pm 0.28$   &    $6.35 \pm 0.29$  \\
F2100W 	& 20.795&	 1208.9	& $5.23 \pm 0.35$   &    $5.33 \pm 0.36$  \\
\enddata
\tablecomments{Dereddened by $A_V=0.85\pm0.17$ mag \citep{Finley26}.}
\end{deluxetable*}

\begin{figure*}[t]
\centering
\figurenum{1}
\includegraphics[width=\linewidth]{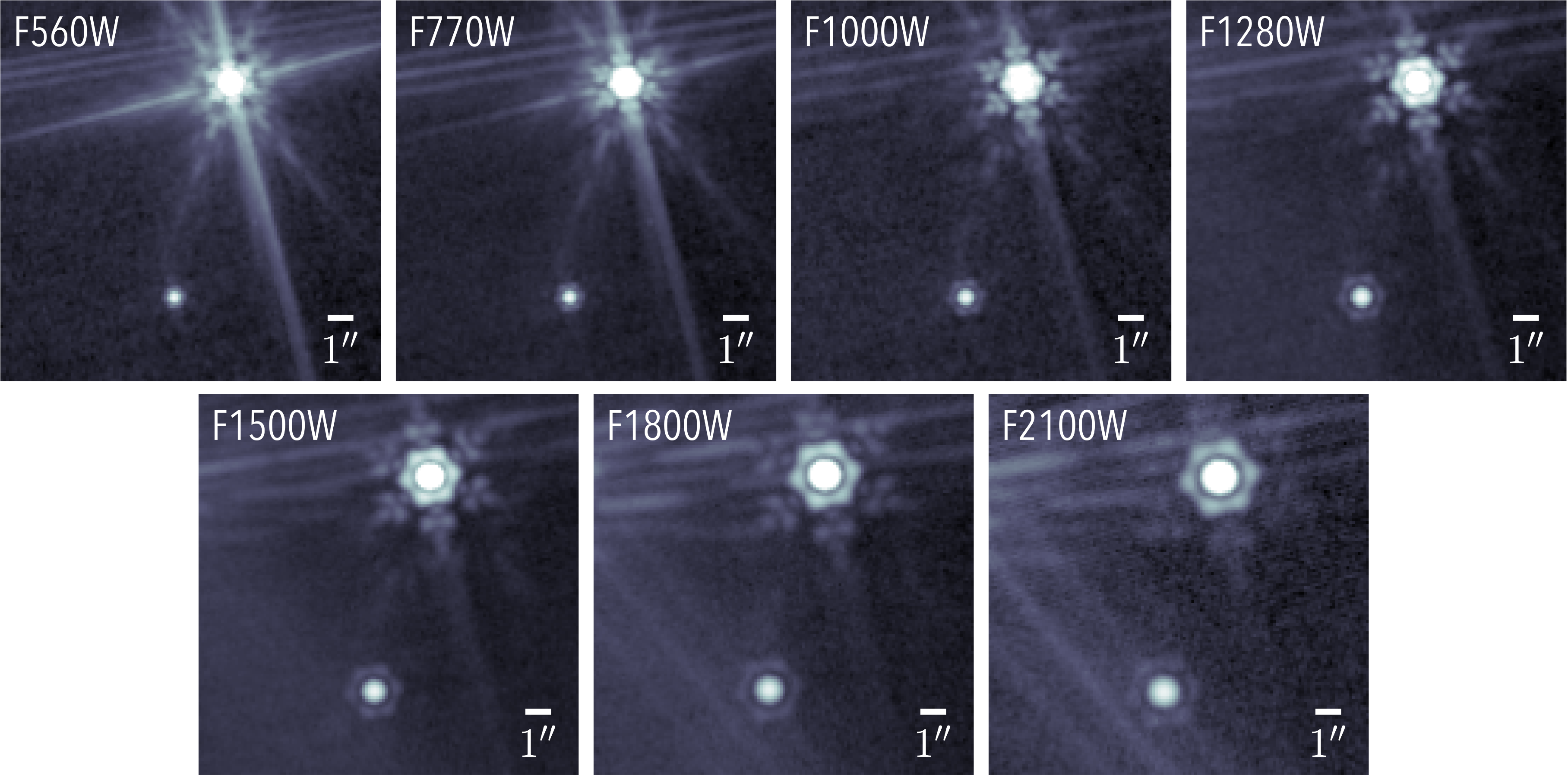}
\caption{JWST MIRI images of the SR 12 system on a logarithmic scale. SR 12 c is located 8\farcs7 southeast of the unresolved binary SR 12 AB. North is up and east is left.}
\label{fig:galery}
\end{figure*}

\section{Analysis}
\subsection{Astrometry}
Figure \ref{fig:galery} displays the SR 12 system in all seven MIRI filters. The host binary SR 12 AB is not spatially resolved. There are no other point sources in the images that could be potential companion candidates. The faint diffraction patterns from two nearby young stellar objects are also present. The closer and brighter source is YLW 13B (2MASS J16272146$-$2441430), which is at 26\farcs9 and a position angle (PA) of 95\fdg7~from SR 12 AB. The diffraction patterns from the more distant source YLW 16A (2MASS J16272802$-$2439335), which is at 172\arcsec~and PA of 42\fdg5~from SR 12 AB, are only visible at $\gtrsim$15~\micron. 

We measure the position of SR 12 c relative to the host binary and compare it with previous measurements. Since F560W is undersampled, astrometry is carried out using the F770W and F1000W images, along with the precise pixel scales and errors ($110.47894\pm0.00036$ mas\,pix$^{-1}$ and $110.40550\pm0.00034$ mas\,pix$^{-1}$, respectively) in \cite{Libralato24}. However, the final astrometric uncertainty is likely to be dominated by any residual image distortion, which we conservatively assume to be 0\farcs03 following the description in the JWST User Documentation. We derive an identical separation of $8\farcs69 \pm 0\farcs03$ and PA of $165\fdg3 \pm 0\fdg2$ for both images. Our measurements agree with the previous results from HST and adaptive optics ($8\farcs67 \pm 0\farcs06$\footnote{SR 12 AB is spatially resolved by HST and ground-based facilities \citep{K11}. The value reported here corresponds to the mean separation between SR 12 A and SR 12 c using Table 3 in \cite{K11}.};  \citealt{K11}) and Spitzer\slash IRAC ($8\farcs62 \pm 0\farcs05$ and $164\fdg8 \pm 0\fdg6$; \citealt{MK22}) within the uncertainties.

\subsection{Photometry}
As SR 12 c is widely separated from the host binary, we do not subtract SR 12 AB in the images before carrying out aperture photometry. We adopt an aperture of theoretical 70\%~encircled energy and the corresponding aperture correction factor in the calibration file \texttt{jwst\_miri\_apcorr\_0014.fits} downloaded from the JWST Calibration References Data System. The adopted aperture radii are 3.87, 4.22, 4.60, 5.18, 5.69, 6.29, and 7.91 pixels for the seven filters, and the corresponding correction factors are 1.44, 1.44, 1.45, 1.48, 1.50, 1.51, and 1.49, respectively. For most filters, this aperture radius is within 1--2 full widths at half maximum. 

Photometric uncertainty is determined by the square root of the quadrature sum of these individual errors: the error map $\sigma_{\rm err}$, the sky emission $\sigma_{\rm sky}$, and systematic errors $\sigma_{\rm sys}$ quantified by \cite{Gordon25}. The error map accounts for detector noise propagation such as Poisson noise, read noise, and other calibration uncertainties. We square the map, sum the pixel values within an aperture centered on the source, and compute the square root of this value to get $\sigma_{\rm err}$.

Sky emission contributes the largest fraction of the photometric error budget. It arises from background fluctuations and contamination from the point-spread functions (PSFs) of nearby stars. To estimate this component, we place 12 independent sky apertures with the same radius as the source aperture around the target. The centers of these apertures are located four aperture radii from the source center. We measure the total counts in each sky aperture and compute their standard deviation, which is adopted as $\sigma_{\rm sky}$. For SR 12 c, the main source of sky emission arises from the PSFs of SR 12 AB at $\lesssim$10\,\micron~and YLW 16A at $\gtrsim$15\,\micron. The scatter is relatively minor, rising from $\lesssim$1\%~of the brightness of SR 12 c at $\lesssim$10\,\micron~to~about 8\%~at 21\,\micron. 

\cite{Gordon25} characterize two systematic errors, $\sigma$(CF) and $\sigma$(repeat), in the absolute flux calibration of MIRI imaging and coronagraphy. $\sigma$(CF) represents the uncertainty in the calibration factor used to convert detector count rates into flux densities. On the other hand, $\sigma$(repeat) is the scatter in the measured fluxes from repeated observations in a given filter. Both systematic errors amount to less than 0.7\%~of the measured source flux density across all seven MIRI filters (Table 8 in \citealt{Gordon25}). Table \ref{tb:phot} also summarizes our MIRI photometry for SR 12 c. SR 12 c is clearly detected in all filters, with signal-to-noise ratios of 112, 120, 120, 54, 38, 22, and 15 in F650W, F770W, F1000W, F1280W, F1500W, F1800W, and F2100W, respectively.

\subsection{SED Fitting with Analytic Disk Models}
With the available multiwavelength data spanning from the UV to the submillimeter, SR 12 c provides a valuable window into the disk around a planetary-mass object. Modeling the SED can constrain the bulk properties of a disk and other microphysics in dust grains (e.g., \citealt{RE21}). Here we use analytic models to constrain the structure of the disk and the dominant heating mechanism. Hydrogen slab modeling of the accretion shock is presented in \cite{Finley26} but not included here because it has a negligible contribution at mid-infrared wavelengths. 

\subsubsection{Two-temperature Blackbody}
Submillimeter detections are rare for planetary-mass companions, so a single blackbody model is typically used to describe the disk's mid-infrared SED. This approach has been applied to the JWST observations of several planetary-mass companions, including GQ Lup B \citep{S21,Cugno24}, YSES-1 b \citep{Hoch25}, and Delorme 1 AB b \citep{Malin25}. For SR 12 c, a single blackbody is insufficient to reproduce the infrared-to-submillimeter SED. We therefore model the disk as a combination of warm and cold components, characterized by blackbody temperatures $T_1$ and $T_2$ and effective radii $r_1$ and $r_2$. The disk flux density $F_\lambda$ is the sum of two scaled blackbodies: 
\begin{equation} \label{eq:1}
F_{\lambda} =\frac{r_1^2}{D^2} \pi B(T_1) + \frac{r_2^2}{D^2} \pi B(T_2),
\end{equation}
where $D$ is the distance to the source, and $B(T)$ is the Planck function. 

\subsubsection{Irradiation}
Since irradiation from the host binary is negligible at such a wide separation, irradiation from SR 12 c itself is expected to play an important, if not dominant, role in the temperature profile of the disk. Regarding the disk geometry, brown dwarf disks have been shown to be less flared than T Tauri disks (e.g., \citealt{Pascucci03, Liu15}), partly due to the lower luminosity of the central objects, which provides less heating, and partly due to their relatively low gas surface density, resulting in weaker coupling between dust and gas and thus more efficient dust settling toward the midplane \citep{DD04}. Since planetary-mass companions are even less luminous, their dust disks may be even flatter than those of brown dwarfs, approaching a flat geometry. 

To model an irradiated flat disk, we adopt the temperature profile in \cite{CG97}: 
\begin{equation} \label{eq:2}
T_{\rm irr}(r) = \bigg(\frac{2}{3\pi}\bigg)^{1/4}\,T_{\rm PMC} \bigg(\frac{r}{R_{\rm PMC}} \bigg)^{-3/4},
\end{equation}
where $T_{\rm PMC}$ and $R_{\rm PMC}$ are the effective temperature and the radius of the companion, respectively. 

To quantify (weak) disk flaring, we consider a flared irradiated disk in \cite{CG97}, whose temperature profile is as follows:
\begin{equation} \label{eq:3}
T_{\rm irr}(r) = \bigg(\frac{\alpha}{2}\bigg)^{1/4}\,T_{\rm PMC} \Bigg(\frac{r} {R_{\rm PMC}}\bigg)^{-1/2},
\end{equation}
where $\alpha$ is the grazing angle. We treat $\alpha$ as a free parameter, so the inferred value is interpreted as an effective, flux-weighted grazing angle. The temperature profile of a flat disk decreases more steeply with radius than that of a flared disk because the grazing angle in a flat disk approximately scales as $r^{-1}$, whereas it has a weaker radial dependence in a flared disk.

\subsubsection{Viscous Heating}
In a differentially rotating disk, viscous heating arises from the dissipation of velocity shear by internal stresses, often parameterized as an effective viscosity (e.g., \citealt{P81}). This dissipation accompanies angular momentum transport that enables accretion, and the released heat reflects the conversion of gravitational potential energy as material spirals inward, potentially contributing to the observed mid-infrared flux.

We follow the temperature profile of the classical model of \cite{SS73}, 
\begin{equation} \label{eq:4}
T_{\rm acc}(r) =  \Bigg[ \frac{3G M_{\rm PMC} \dot{M}_{\rm PMC}}{8\pi \sigma r^3} \Bigg(1-\sqrt{\frac{r_{\rm in}}{r}}\Bigg)\Bigg]^{1/4} ,
\end{equation}
where $r_{\rm in}$ is the inner-disk radius, $M_{\rm PMC}$ is the mass of the companion, and $\dot{M}_{\rm PMC}$ is the accretion rate.

\subsubsection{The Full Disk Profile}
The total disk emission depends on the combination of the heating mechanisms described above. For instance, in a composite model including both irradiation and viscous heating, the temperature profile is given by
\begin{equation} \label{eq:5}
T(r)^4 = T_{\rm irr}(r)^4 + T_{\rm acc}(r)^4. 
\end{equation}
The spectral flux density from the disk is computed by integrating the Planck function $B_\lambda(T(r))$ over a logarithmically spaced radial grid extending from the inner radius $r_{\rm in}$ to the outer radius $r_{\rm out}$. This logarithmic spacing provides finer sampling of the inner disk, where the physical conditions vary more rapidly. The resulting flux is scaled by the disk inclination $i$:
\begin{equation} \label{eq:6}
F_{\rm disk} = \int^{r_{\rm out}}_{r_{\rm in}} \frac{2\pi r}{D^2} B_\lambda (T(r)) \cos i \, dr. 
\end{equation}
The total flux density $F_{\rm full}$ is obtained by adding the photosphere to the disk emission,  
\begin{equation} \label{eq:7}
F_{\rm full} =  F_{\rm phot} + F_{\rm disk}.
\end{equation}
We adopt the 2600 K log $g$ = 4.0 BT-Settl model \citep{Allard12} for the photosphere of SR 12 c.

\subsubsection{Bayesian SED Fitting}
The fitting procedure and parameter inference are based on Markov Chain Monte Carlo (MCMC). We use a custom implementation of the Metropolis--Hastings algorithm to generate posterior samples. For each model evaluation, the likelihood $\mathcal{L}$ is computed in logarithmic flux space:
\begin{equation} \label{eq:8}
\ln \mathcal{L} = -\frac{1}{2}\sum_k 
\left[
\frac{(\Delta_k)^2}{\sigma_k^2}
+ \ln(2\pi\sigma_k^2)
\right],
\end{equation}
where $\Delta_k = \log F_{{\rm obs},k} - \log F_{{\rm model},k}$ is the residual between the observed and modeled logarithmic flux at wavelength $k$, and $\sigma_k$ is the corresponding uncertainty in logarithmic space, obtained by combining the observational uncertainties with model uncertainties propagated from $T_{\rm PMC}$, $R_{\rm PMC}$, and $D$. 

Because only broadband photometry is available, without resolved imaging or kinematic constraints from spectroscopy, the disk inclination is expected to be only weakly constrained and may introduce additional degeneracies among the model parameters. We therefore adopt a representative intermediate inclination of $45\degr$ for the main analysis. An alternative fit in which the inclination is treated as a free parameter is presented in the \hyperref[Appendix]{Appendix} and yields qualitatively consistent results regarding the preferred disk structure and heating mechanism.

We assume log-uniform priors for $r_{\rm in}$, $r_{\rm out}$, and $\dot{M}$, reflecting our limited knowledge of their characteristic scales. For $\alpha$, we adopt uniform priors over 0--1 rad. During the burn-in phase (50\% of the chain), adaptive step sizes are employed to achieve an acceptance rate of 0.3. After burn-in, the step sizes are held fixed. The posterior distributions are produced by combining the chains after removing the burn-in. Convergence is evaluated using the Gelman--Rubin statistic ($\hat{R}<1.01$) together with the effective sample size, requiring more than 1000 independent posterior samples. Due to parameter degeneracies, the maximum a posteriori solution does not necessarily coincide with the peaks of the marginal posterior distributions. We therefore adopt the median values and the 68\%~credible intervals of the marginalized posterior distributions as representative parameter estimates. 

Model comparison is conducted using the Bayesian information criterion (BIC; \citealt{Schwarz78}), defined as 
\begin{equation} \label{eq:9}
\mathrm{BIC}  = -2 \ln \mathcal{L_{\rm max}} + N_{\rm param} \ln N_{\rm data}, 
\end{equation}
where $\mathcal{L_{\rm max}}$ is the maximized likelihood of the model, $N_{\rm param}$ is the number of model parameters, and $N_{\rm data}$ is the number of photometric points. A model with a substantially lower BIC value is better supported by the data, whereas models with similar BIC values cannot be distinguished based on the available data. Following \cite{Raftery95}, differences in BIC ($\Delta$BIC) between models were interpreted as follows: 0--2, 2--6, 6--10, and $>$10, representing negligible, weak, moderate, and strong evidence, respectively, in favor of the model with the lower BIC.

\begin{deluxetable}{p{0.75\linewidth}l p{0.75\linewidth}l}
\tablecaption{Best-fit Model Parameters\label{tab:best_params}}
\tablewidth{\columnwidth}
\tablehead{Parameter & Value}
\startdata
\multicolumn{2}{c}{Two-temperature Blackbody} \\
\hline
$r_1$ [$R_{\rm Jup}$] 		& 	$17.0^{+1.4}_{-1.3}$ \\
$r_2$ [$R_{\rm Jup}$] 		& 	$766^{+61}_{-57}$ \\
$T_1$ [K]					& 	$533^{+27}_{-25}$ 	\\
$T_2$ [K]					& 	$80.0^{+2.8}_{-2.9}$ \\
$L_{\rm disk}/L_{\rm SR12c}$ 	& 	$0.28^{+0.08}_{-0.06}$	\\
BIC 						&	$-25.8$	\\
\hline
\multicolumn{2}{c}{Irradiated Flat Disk} \\
\hline
$r_{\rm in}$ [$R_{\rm Jup}$] 	& 	$2.5^{+0.8}_{-0.4}$	\\
$r_{\rm out}$ [$R_{\rm Jup}$] 	& 	$5500^{+4900}_{-2200}$ 	\\
$L_{\rm disk}/L_{\rm SR12c}$	& 	$0.22^{+0.08}_{-0.07}$	\\
BIC 						&	$-11.5$				\\
\hline
\multicolumn{2}{c}{Irradiated Flared Disk} \\
\hline
$r_{\rm in}$ [$R_{\rm Jup}$] 	& 	$2.4^{+0.9}_{-0.4}$	\\
$r_{\rm out}$ [$R_{\rm Jup}$] 	& 	$1090^{+140}_{-120}$		\\
$\alpha$ 					& 	$0.05^{+0.01}_{-0.01}$ 	\\
$L_{\rm disk}/L_{\rm SR12c}$	& 	$0.23^{+0.07}_{-0.05}$ 	\\
BIC 						&	$-2.4$				\\
\hline
\multicolumn{2}{c}{Viscous-heating-only Disk} \\
\hline
$r_{\rm in}$ [$R_{\rm Jup}$] 	& 	$2.1^{+0.2}_{-0.1}$ 			\\
$r_{\rm out}$ [$R_{\rm Jup}$] 	& 	$2800^{+610}_{-460}$ 			\\
$\log \dot{M}$ [$M_{\odot}$ yr$^{-1}$] 	& 	$-9.75^{+0.04}_{-0.03}$ 	\\
$L_{\rm disk}/L_{\rm SR12c}$	& 	$0.19^{+0.05}_{-0.04}$ \\
BIC 						&	$-24.6$	\\
\hline
\multicolumn{2}{c}{Irradiated Flat Disk with Viscous Heating} \\
\hline
$r_{\rm in}$ [$R_{\rm Jup}$] 	&	$2.5^{+0.8}_{-0.4}$  		\\
$r_{\rm out}$ [$R_{\rm Jup}$] 	& 	$5500^{+4900}_{-2200}$ 	\\
$\log \dot{M}$ [$M_{\odot}$ yr$^{-1}$] 		& 	$-13.7^{+0.5}_{-0.2}$ \\
$L_{\rm disk}/L_{\rm SR12c}$	& 	$0.22^{+0.08}_{-0.06}$	\\
BIC 						&	$-14.0$	\\
\enddata
\tablecomments{Except for the two-temperature blackbody model, which has no inclination dependence, the disk inclination is fixed at $45\degr$ for all other models.}
\end{deluxetable}

\begin{figure*}[!t]
\centering
\figurenum{2}
\includegraphics[width=\linewidth]{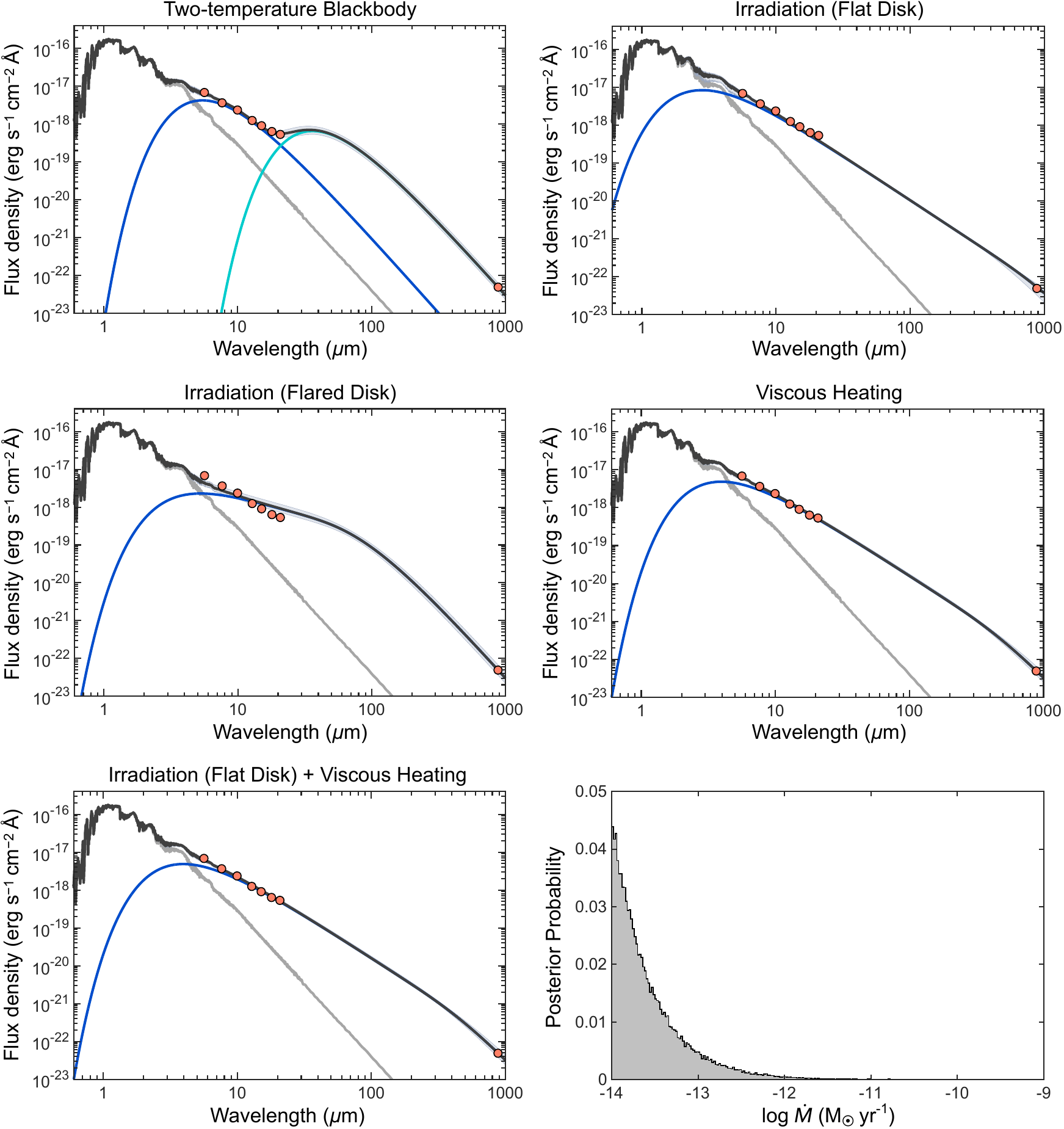}
\caption{SED fitting of SR 12 c with analytic models. The best-fit model (posterior median) is in blue, the photosphere is in gray, and their sum is in black. The gray shaded band is the 95\%~posterior predictive interval. The JWST and ALMA measurements are shown in orange. The SR 12 c disk is well described by a two-temperature blackbody (upper left), as well as the irradiated flat-disk model (upper right). The irradiated flared-disk model (middle left) overpredicts the disk flux at $\gtrsim$10~\micron~because the outer regions would intercept more energy. The viscous-heating-only model (middle right) can reproduce the data, but it requires an accretion rate that is much higher than the observed value of $\sim$$10^{-11}~M_\odot$ yr$^{-1}$. The composite model of irradiation and viscous heating can reproduce the overall SED (bottom left), but the accretion rate is not meaningfully constrained because its posterior distribution accumulates at the lower prior bound (bottom right). This suggests that the observed SED is governed by irradiation, with viscous heating playing a minor role.}
\label{fig:SR12c_SED}
\end{figure*}

\section{Results and Discussion}  
\label{sect:SED_best_models}
Here we present the results of our SED fitting and discuss their implications. To minimize potential systematic effects associated with combining observations from widely separated epochs, we fit our disk models only to the JWST and ALMA observations, which were obtained at similar epochs and provide higher signal-to-noise ratios than other near-infrared measurements in the literature. Table \ref{tab:best_params} summarizes our MCMC analysis, and Figure \ref{fig:SR12c_SED} illustrates the best-fit models. These simple models capture the basic physics and provide global constraints on the disk geometry and thermal structure. 

\subsection{Two-component Blackbody}
The SED of SR 12 c is well described by a two-component blackbody model, yielding distinct characteristic temperatures and effective radii for the warm and cold components. Adopting a graybody prescription would yield an even larger effective radius for the cooler component. However, the interpretation of these multitemperature components is degenerate because the disk structure and dust properties are poorly constrained. For example, grain growth and dust settling can modify the dust opacity and vertical distribution, while variations in irradiation geometry, such as inner-disk shadowing, can produce strong radial temperature gradients within an otherwise continuous disk. Resolved ALMA observations of protoplanetary disks have begun to reveal the radial and vertical distributions of dust and gas (e.g., \citealt{Kurtovic21,Villenave25}), whereas such spatially resolved constraints remain unavailable for disks around planetary-mass objects. Consequently, the multitemperature SED should be regarded as suggestive of radial and\slash or vertical variations in the disk structure and dust properties, rather than as direct evidence for discrete radial substructures.

\subsection{A Flat Dust Disk with Minor Flaring}
As illustrated in Figure \ref{fig:SR12c_SED}, the irradiated flat-disk model provides a significantly better fit than the irradiated flared-disk model. The observed SED shows a relatively steep slope in the mid-infrared. In contrast, flared disks intercept more stellar radiation at large radii and therefore tend to overpredict the mid- to far-infrared emission. A flat disk, on the other hand, naturally suppresses the energy inception in the outer disk, reducing the observed flux at long wavelengths. The substantial $\Delta$BIC = 9.1 also indicates that the current SED does not require strong disk flaring. 

The outer disk radius $r_{\rm out}$, however, is not strongly constrained by the current data. Using the flat-disk model, the posterior distribution of $r_{\rm out}$ is broad, yielding $r_{\rm out} = 2.6^{+2.4}_{-1.0}$ au. For the flared-disk model, we obtain $r_{\rm out} = 0.52^{+0.07}_{-0.06}$ au. Both estimates are broadly consistent with the ALMA dust continuum radius of 0.5--7 au (adjusted by a larger distance) in \cite{W22}. Mid-infrared data are intrinsically less sensitive to the disk extent; therefore, far-infrared or resolved submillimeter imaging will be required to place tighter constrains on $r_{\rm out}$.

\subsection{The Minor Role of Viscous Heating}
We investigate the relative importance of irradiation and viscous heating by comparing the irradiation-only model (flat disk), the viscous-heating-only model, and the combined model including both heating mechanisms. The viscous-heating-only model can reproduce the observed SED and yields a lower BIC than the irradiation-only model. However, the accretion rate required by this model, $10^{-9.8}~M_\odot$ yr$^{-1}$, is $\sim$20 times higher than estimates obtained from independent diagnostics, including the UV-optical continuum and emission-line tracers ($10^{-11.1}~M_\odot$ yr$^{-1}$; \citealt{SM19,Finley26}). This result does not by itself demonstrate that viscous heating is subdominant. Instead, it highlights the degeneracy of SED modeling: a purely viscous model can reproduce the observed SED but requires a substantially higher accretion rate than independently inferred.

The contribution of viscous heating is further constrained in the bottom panel of Figure \ref{fig:SR12c_SED}, where we consider the composite model of irradiation and viscous heating. To account for potential accretion variability, we adopt a broad prior on the accretion rate, ranging from $10^{-14}$ to $10^{-9}~M_\odot$ yr$^{-1}$. The posterior accumulates at the lower prior bound, and this behavior persists when the lower bound is further decreased. This indicates that the current data do not provide a meaningful lower bound on the accretion rate, suggesting that irradiation alone can largely account for the observed SED.

As a direct test of this interpretation, we adopt an inverse approach by fixing the accretion rate to the average value of $10^{-11.08\pm0.40}\,M_{\odot}\, \mbox{yr}^{-1}$ reported by \cite{SM19} in both the viscous-heating-only and composite models. With the accretion rate fixed, the viscous-heating-only model underpredicts the observed flux densities by a factor of a few and fails to reproduce the SED. This demonstrates that the independently inferred accretion rate is insufficient for viscous dissipation alone to heat the disk to the temperatures required to produce the observed emission. In contrast, the composite model provides a good fit to the data. The inferred disk parameters and BIC ($r_{\rm in} = 2.5^{+0.8}_{-0.5}$ au, $r_{\rm out} = 5100^{+4700}_{-1900}$ au, $L_{\rm disk}/L_{\rm SR12c} = 0.23^{+0.08}_{-0.06}$, BIC = $-$12.4) are nearly identical to those obtained when the accretion rate is left free ($r_{\rm in} = 2.5^{+0.8}_{-0.4}$ au, $r_{\rm out} = 5500^{+4900}_{-2200}$ au, $L_{\rm disk}/L_{\rm SR12c} = 0.22^{+0.08}_{-0.06}$, BIC = $-$14.0; see Table \ref{tab:best_params}). This indicates that, once irradiation is included, the inferred disk parameters are largely insensitive to the accretion rate, because the disk thermal structure is primarily determined by irradiation rather than viscous dissipation.

\begin{figure}[!h]
\centering
\figurenum{3}
\includegraphics[width=\linewidth]{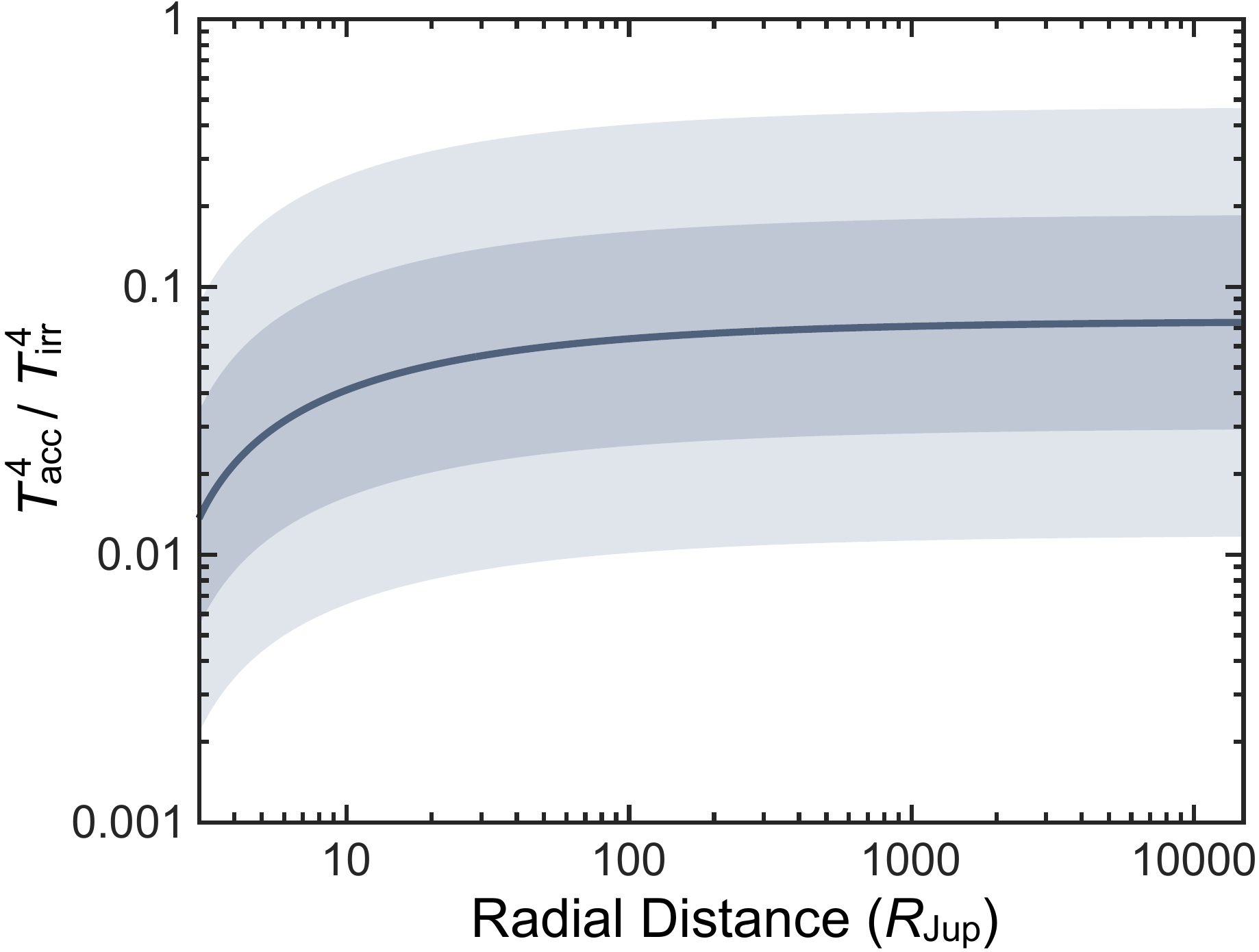}
\caption{The energy flux ratio between irradiation and viscous heating, calculated using Equations (\ref{eq:2}) and (\ref{eq:4}). The light and dark gray bands are the 68\%~and~95\%~credible intervals, respectively.}
\label{fig:Fratio}
\end{figure}

Figure \ref{fig:Fratio} shows the energy flux ratio $L_{\rm acc}/L_{\rm irr}=T_{\rm acc}^4/T_{\rm irr}^4$ between viscous heating and irradiation using the average accretion rate in \cite{SM19}. The ratio is $\sim$0.07 over most of the disk and even lower in the inner regions, indicating that irradiation dominates the thermal heating throughout the disk. We note, however, that the precise magnitude of the viscous contribution remains uncertain because line-luminosity-based accretion estimates may underestimate the true accretion rate if a significant fraction of the accretion energy is emitted in Ly$\alpha$ \citep{Aoyama21}. 

Including viscous heating nonetheless provides a modest improvement in the statistical fit. The composite model has a lower BIC than the irradiated flat-disk model ($\Delta$BIC = 2.5; Table \ref{tab:best_params}), suggesting that the data may weakly favor a small additional contribution from viscous heating. Taken together, these results indicate that SED fitting alone does not uniquely constrain the accretion rate due to model degeneracy. However, when the accretion rate is constrained by independent diagnostics, viscous heating alone is insufficient to reproduce the observed SED, whereas irradiation readily does so. We therefore conclude that irradiation is the dominant heating mechanism of the disk, while viscous heating provides at most a subdominant contribution.

\subsection{The Fractional Disk Luminosity}
The fractional disk luminosity, $L_{\rm disk}/L_\star$, reflects the fraction of stellar radiation reprocessed by the disk and thus serves as a diagnostic of its evolutionary stage. We calculate $L_{\rm disk}/L_{\rm SR12c}$ by integrating the best-fit disk model and comparing it to SR 12 c's bolometric luminosity. Table \ref{tab:best_params} lists these ratios for all models, which are 0.2--0.3. These values are broadly consistent with those of T Tauri stars (0.1--0.2; \citealt{Oliveira13}) and GQ Lup B ($\sim$0.1; \citealt{Cugno24}), but significantly higher than those of more evolved disks (e.g., $10^{-6}$--$10^{-2}$ for debris disks; \citealt{KM14}).

\subsection{The Low 10~\micron~Excess}
\label{sect:10um_excess}
Amorphous and crystalline silicate grains in optically thin disk atmospheres show prominent mid-infrared spectral features (e.g., \citealt{Natta00}). The dominant broad feature at $\sim$9.7~\micron~is commonly attributed to the Si--O--Si stretching mode in amorphous olivine- and pyroxene-like silicates. Narrower features associated with crystalline species include a characteristic band at $\sim$10.5--10.7~\micron~from enstatite (MgSiO${}_3$) and a band at $\sim$11.3~\micron~from forsterite (Mg${}_2$SiO${}_4$). Together, these features form a wide continuum excess, sometimes with several peaks, around 10~\micron. This 10~\micron~feature is commonly observed for disks around young stars (e.g., \citealt{Natta00,vB05,Olofsson09}) and brown dwarfs (e.g., \citealt{Apai05,Riaz09}). Its shape and strength provide useful diagnostics of dust processing, including grain growth, crystallization, and dust settling (e.g., \citealt{Jang25}).

\begin{figure}[t]
\centering
\figurenum{4}
\includegraphics[width=\linewidth]{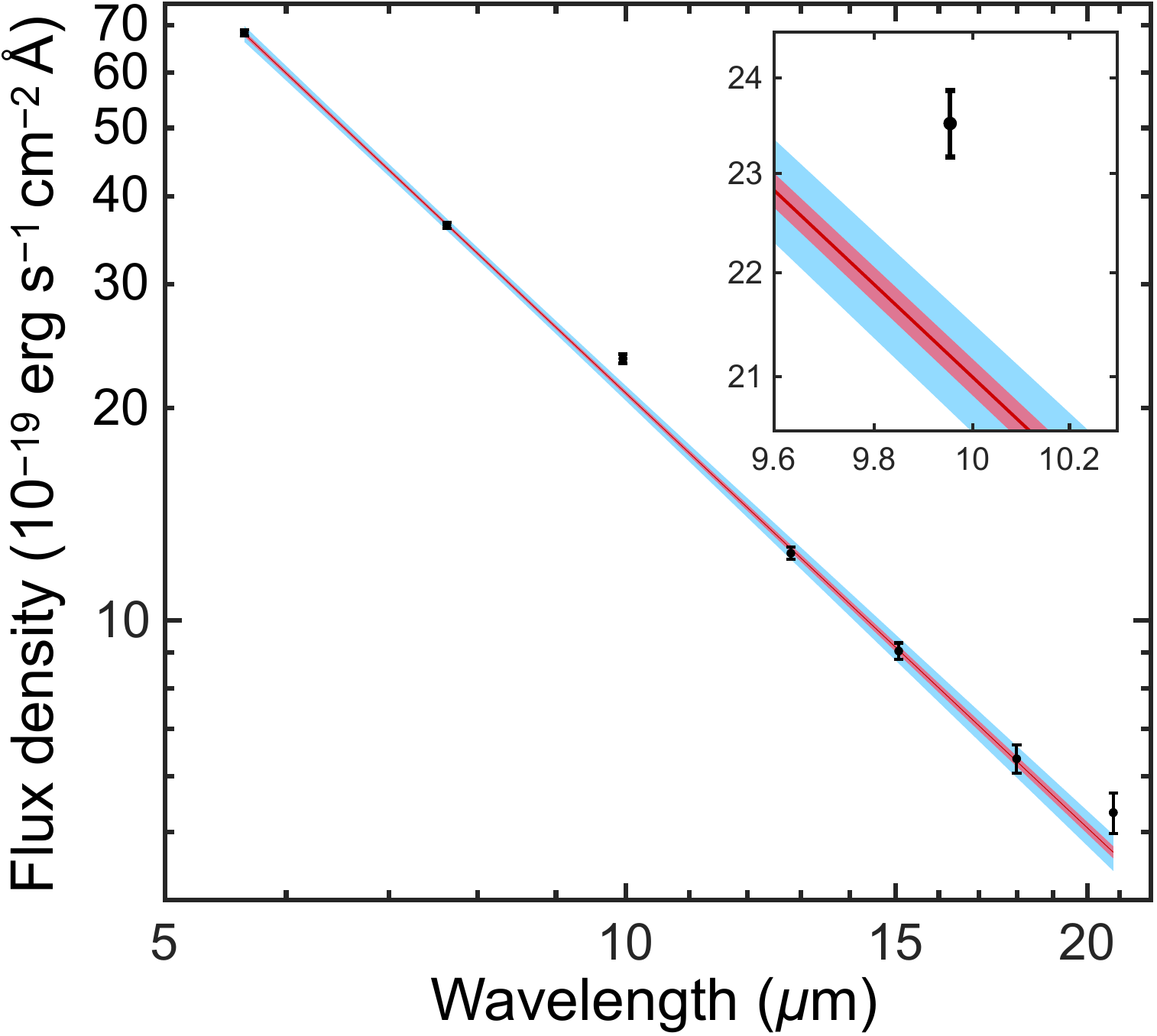}
\caption{The weak 10~\micron~excess of SR 12 c. We perform a linear regression on the MIRI data points, excluding the F1000W flux density. The red and blue belts represent the 68\%~and 95\%~credible intervals, respectively. The F1000W measurement is 1.1 times higher than the predicted continuum (2.35 versus $2.12\times10^{-18}$ erg s$^{-1}$ cm$^{-2}$ \AA), indicating a weak excess of silicate emission.}
\label{fig:SR12c_10um}
\end{figure}

We investigate if SR 12 c exhibits a strong 10~\micron~excess or absorption by performing linear regression to the dereddened MIRI flux densities except the F1000W point. In Figure \ref{fig:SR12c_10um}, we plot the best-fit line, log $F_\lambda$ = $(3.37\pm0.02) - (2.05\pm0.02)$ log $\lambda$, and the 68\%~and 95\%~credible intervals. Our F1000W photometry is 11\%~above the fitted continuum, indicating a weak silicate excess. Owing to the small photometric uncertainty, this excess is detected at 6.0$\sigma$ significance. For other planetary-mass companions, recent MIRI spectroscopy of YSES-1 b has revealed the presence of submicron olivine grains in its disk \citep{Hoch25}, while no silicate features were detected for GQ Lup B \citep{Cugno24} and Delorme 1 AB b \citep{Malin25}. This suggests that the presence of a 10~\micron~silicate feature is not solely determined by the age of the system, as objects spanning a broad range of ages show different silicate properties---GQ Lup B (1--3 Myr) $<$ SR 12 c (3--4 Myr) $<$ YSES-1 b (17--27 Myr; \citealt{Darcis26}) $<$ Delorme 1 AB b (30--45 Myr)---but is likely regulated by dust processing mechanisms and possibly the replenishment of small grains through planetesimal collisions, as proposed for YSES-1 b by \cite{Hoch25}.

\cite{K07} studied the relationship between stellar irradiation and the intensity of the 10~\micron~feature. They showed that the distance to the silicate emission zone follows $R=0.35\, (L/L_{\odot} )^{0.56}$ au, implying that the emission region lies much closer to planetary-mass objects than to stars. For SR 12 c, we derive a distance of $9\times10^{-3}$ au, or 19 $R_{\rm Jup}$, suggesting that most of the F1000W excess originates from the innermost region of the disk. The disk temperature at this location, using Equation (\ref{eq:2}), is approximately 310 K, which is well below the $\gtrsim$800 K required to anneal amorphous silicates (e.g., \citealt{Gail98}). If the SR 12 c disk is not sufficiently warm to anneal a substantial fraction of grains, the F1000W excess is likely dominated by emission from amorphous silicates (olivine and pyroxene). Alternatively, if the excess is instead due to crystalline grains, amorphous silicates may be annealed at radii $\lesssim$4.5 $R_{\rm Jup}$ and subsequently transported outward. 

Grain growth, dust settling, and disk inclination may contribute to the low F1000W excess. Submicron-sized grains exhibit strong resonances in the Si--O--Si molecular bonds, producing prominent spectral features, while grains larger than a few microns progressively broaden, flatten, and weaken the silicate feature due to their more wavelength-independent opacity (e.g., \citealt{Bouwman08}). Dust settling further reduces the population of small, optically thin grains in the disk atmosphere, thereby weakening the observed mid-infrared silicate emission (e.g., \citealt{DD08}). Alternatively, a highly inclined disk can increase the optical depth of the disk surface layer, reducing the contrast between the silicate feature and the underlying photospheric continuum. At high inclinations, the cooler outer disk may also absorb dust emission from the warmer inner disk. However, a highly inclined geometry is difficult to reconcile with the low-to-moderate inclinations inferred from our SED fitting (see the \hyperref[Appendix]{Appendix}) and the low extinction \citep{SM18,Finley26}. Therefore, the observed weak 10~\micron~excess implies that SR 12 c hosts a young disk with evolved dust properties. Without spatially resolved observations, the degeneracy among grain growth, dust settling, and disk inclination remains significant. Future MIRI spectroscopy and multiband ALMA observations may help to partially break the degeneracy and shed light on the physical origin of the weak excess.

\subsection{Formation of SR 12 c}
Different formation pathways of wide-orbit planetary-mass companions may leave imprints on their surrounding disks. Proposed formation scenarios include gravitational instability in a massive circumstellar disk \citep{Boss97}, core accretion with subsequent dynamical scattering (e.g., \citealt{RF96,Veras09}), capture of free-floating objects \citep{PK12}, and cloud fragmentation (e.g., \citealt{Bate09}).

These scenarios can be constrained by the observed disk properties. Disk fragmentation requires a massive and extended disk capable of becoming gravitationally unstable. However, current infrared and millimeter observations do not show evidence for such a disk reservoir around SR 12 AB (\citealt{MK22,W22,Finley26}). Both scattering and capture involve strong dynamical interactions, which may truncate or disrupt the disk and suppress ongoing accretion. Moreover, high-contrast imaging campaigns have not detected inner massive companions capable of inducing such scattering events (e.g., \citealt{Bryan16,Pearce19}).

Alternatively, cloud fragmentation is widely considered the dominant pathway for low-mass star formation and may also contribute to brown dwarf formation. SR 12 c is broadly consistent with several star--disk correlations observed among young stellar and substellar populations, including the millimeter flux density--stellar mass relation \citep{W22} and the accretion rate--stellar mass relation \citep{Finley26}. The disk radius of SR 12 c derived in this work also lies within the scatter of the disk size--millimeter flux density relation in young star-forming regions (see Figure 6 in \citealt{Hendler20}). The consistency of SR 12 c with star--disk scaling relations suggests that its formation pathway is more similar to that of low-mass stars and brown dwarfs.

\section{Conclusion}
Circumplanetary disks regulate the formation and evolution of giant planets and their satellites, playing a pivotal role in understanding planet formation and the assembly of moons. SR 12 c is the first wide-orbit planetary-mass companion whose disk has been directly imaged with ALMA. This provides a unique opportunity to constrain the disk's radial extent, substructures, heating mechanisms, and silicate features, particularly through infrared observations.

We present JWST/MIRI broadband imaging of the SR 12 system from 5.6 to 21~\micron. The companion's SED from mid-infrared to submillimeter wavelengths is analyzed using analytic disk models in an MCMC approach. Phenomenologically, the disk SED can be described by a two-component blackbody model with a large contrast in characteristic temperatures and radii. However, this apparent two-temperature structure does not necessarily imply the presence of a physical gap in the disk.

We then consider physically motivated models incorporating different disk geometries and heating mechanisms. We find that the SED of the dust disk is more consistent with a geometrically flat disk with little flaring. The dust disk appears compact; the inner-disk radius is very close to SR 12 c ($\sim$3 $R_{\rm Jup}$), while the outer disk radius has a median of 2.6 au, with a broad posterior distribution spanning $\sim$2--5 au. This reflects the fact that mid-infrared measurements are less sensitive to the disk extent. 

We demonstrate that irradiation from SR 12 c is dominant over viscous heating in the disk's thermal budget. The viscous-heating-only model requires a high accretion rate that is not fully compatible with the observations. In the composite model of both irradiation and viscous heating, the posterior distribution of the accretion rate consistently piles up at the lower bound of the prior, indicating that the observed SED can be largely explained by irradiation. The fractional disk luminosity (0.2--0.3) is comparable to that of T Tauri disks. 

We find a weak F1000W excess, only about 11\% above the local continuum. This weak silicate emission suggests evolved dust grains, with grain growth and dust settling likely weakening the emission feature. Finally, we find that SR 12 c is broadly consistent with the observed star--disk scaling relations, suggesting that its formation may share similarities with that of low-mass stars and brown dwarfs.

\begin{acknowledgments}
We thank the reviewer for the constructive comments and suggestions, which have significantly improved the manuscript. Y.-L.W. is grateful for the support from the National Science and Technology Council (113--2636--M--003--001--). B.P.B. acknowledges support from the Alfred P. Sloan Foundation. G.J.H. acknowledges support from National Natural Science Foundation of China general program 12573031 and grant IS23020 from the Beijing Natural Science Foundation. This work is based on observations made with the NASA\slash ESA\slash CSA James Webb Space Telescope. The data were obtained from the Mikulski Archive for Space Telescopes at the Space Telescope Science Institute, which is operated by the Association of Universities for Research in Astronomy, Inc., under NASA contract NAS 5-03127 for JWST. These observations are associated with program GO 2311. 
\end{acknowledgments}

\facility{JWST (MIRI).}
\software{MATLAB.}

\appendix
\phantomsection \label{Appendix}

\begin{deluxetable}{p{0.75\linewidth}l p{0.75\linewidth}l}
\tablewidth{\columnwidth}
\tablenum{A1}
\tablecaption{Best-fit Model Parameters \label{tab:best_params_free_incl}}
\tablehead{Parameter & Value}
\startdata
\multicolumn{2}{c}{Irradiated Flat Disk} \\
\hline
$r_{\rm in}$ [$R_{\rm Jup}$] 	& 	$2.9^{+1.2}_{-0.7}$		\\
$r_{\rm out}$ [$R_{\rm Jup}$] 	& 	$4300^{+4700}_{-1700}$ 	\\
$i$ [$\degr$] 				& 	$34^{+10}_{-13}$		\\
$L_{\rm disk}/L_{\rm SR12c}$	& 	$0.22^{+0.09}_{-0.06}$	\\
BIC 						&	$-9.7$	\\
\hline
\multicolumn{2}{c}{Irradiated Flared Disk} \\
\hline
$r_{\rm in}$ [$R_{\rm Jup}$] 	& 	$2.8^{+1.9}_{-0.7}$		\\
$r_{\rm out}$ [$R_{\rm Jup}$] 	& 	$3300^{+1300}_{-1300}$	\\
$i$ [$\degr$] 				& 	$85^{+2}_{-8}$ 			\\
$\alpha$ 					& 	$0.45^{+0.35}_{-0.28}$ 	\\
$L_{\rm disk}/L_{\rm SR12c}$	& 	$0.24^{+0.07}_{-0.06}$ 	\\
BIC 						&	$-3.5$	\\
\hline
\multicolumn{2}{c}{Viscous-heating-only Disk} \\
\hline
$r_{\rm in}$ [$R_{\rm Jup}$] 	& 	$2.3^{+0.6}_{-0.3}$ 		\\
$r_{\rm out}$ [$R_{\rm Jup}$] 	& 	$3600^{+2100}_{-1100}$ 	\\
$i$ [$\degr$] 				& 	$60^{+15}_{-22}$ 	\\
$\log \dot{M}$ [$M_{\odot}$ yr$^{-1}$] 		& 	$-9.5^{+0.4}_{-0.3}$ 	\\
$L_{\rm disk}/L_{\rm SR12c}$	& 	$0.20^{+0.06}_{-0.05}$ \\
BIC 						&	$-20.0$	\\
\hline
\multicolumn{2}{c}{Irradiated Flat Disk with Viscous Heating} \\
\hline
$r_{\rm in}$ [$R_{\rm Jup}$] 	&	$2.9^{+1.2}_{-0.7}$  		\\
$r_{\rm out}$ [$R_{\rm Jup}$] 	& 	$4300^{+4700}_{-1700}$ 	\\
$i$ [$\degr$] 				& 	$35^{+10}_{-13}$ 		\\
$\log \dot{M}$ [$M_{\odot}$ yr$^{-1}$] 		& 	$-13.7^{+0.5}_{-0.2}$ \\
$L_{\rm disk}/L_{\rm SR12c}$	& 	$0.22^{+0.09}_{-0.07}$	\\
BIC 						&	$-14.2$	\\
\enddata
\end{deluxetable}

Here we present the SED fitting results with the disk inclination $i$ treated as a free parameter. We assume an isotropic orientation, corresponding to a prior proportional to $\sin i$. Table \ref{tab:best_params_free_incl} lists the best-fit parameters. The resulting fit is qualitatively consistent with that presented in the main text. The inferred disk remains compact, flat, and irradiation-dominated, suggesting that our main conclusions are largely insensitive to the treatment of the disk inclination.

\end{document}